%% file: 0-paper.tex
\documentclass[sigconf]{acmart}
\usepackage{fancyhdr}
\usepackage{graphicx}
\usepackage{tabularx}
\usepackage{subcaption}
\usepackage{textcomp}
\usepackage{siunitx}
\usepackage{amsmath, amsthm}
\usepackage{graphbox}
\usepackage{multirow}
\usepackage{makecell}
\usepackage{fixltx2e}
\usepackage{setspace}
\usepackage{booktabs}
\usepackage{array}
\usepackage{tabularx}
\usepackage{calc}
\usepackage{makecell}
\usepackage[T1]{fontenc}
\usepackage[utf8]{inputenc}
\usepackage[font=small,labelfont=bf]{caption}
\usepackage{booktabs}
\usepackage{subcaption}
\usepackage{enumitem}
\usepackage{multirow}
\usepackage{threeparttable}
\usepackage{makecell}
\usepackage{hhline}

\AtBeginDocument{%
	\providecommand\BibTeX{{%
			\normalfont B\kern-0.5em{\scshape i\kern-0.25em b}\kern-0.8em\TeX}}}

\setcopyright{acmcopyright}
\copyrightyear{2018}
\acmYear{2018}
\acmDOI{XXXXXXX.XXXXXXX}

\acmConference[Conference acronym 'XX]{Make sure to enter the correct
	conference title from your rights confirmation emai}{June 03--05,
	2018}{Woodstock, NY}
\acmPrice{15.00}
\acmISBN{978-1-4503-XXXX-X/18/06}

\begin{document}
	\begin{sloppypar}
		
	\title{When Fine-Tuning is Not Enough: Lessons from HSAD on Hybrid and Adversarial Audio Spoof Detection}

		\author{Bin Hu$^{1}$, Kunyang Huang$^{2}$, Daehan Kwak$^{1}$, Meng Xu$^{1}$, and Kuan Huang$^{1}$}
	\affiliation{
		\institution{$^{1}$ Department of Computer Science and Technology, Kean University, USA \\
    $^{2}$ Department of Computer Science and Technology, Wenzhou-Kean University, China}
		\city{}
		\state{}
		\country{}}
		\renewcommand{\shortauthors}{Bin et al.}
	\input{abstract}


\maketitle
    \input{01-introduction}
    \input{02-Relatedwork}
    \input{3-dataset-detection}
	\input{03-Datasets}

    \input{04-Methodology}
	\input{05-Experiment}

    \input{06-Conclusion}
    \input{07-Acknowledgement}



	\appendix
	
\end{sloppypar}

\end{document}

%% file: abstract.tex
\begin{abstract}
The rapid advancement of AI has enabled highly realistic speech synthesis and voice cloning, posing serious risks to voice authentication, smart assistants, and telecom security. While most prior work frames spoof detection as a binary task, real-world attacks often involve hybrid utterances that mix genuine and synthetic speech, making detection substantially more challenging. To address this gap, we introduce the Hybrid Spoofed Audio Dataset (HSAD), a benchmark containing 1,248 clean and 41,044 degraded utterances across four classes: human, cloned, zero-shot AI-generated, and hybrid audio. Each sample is annotated with spoofing method, speaker identity, and degradation metadata to enable fine-grained analysis. We evaluate six transformer-based models, including spectrogram encoders (MIT-AST, MattyB95-AST) and self-supervised waveform models (Wav2Vec2, HuBERT). Results reveal critical lessons: pretrained models overgeneralize and collapse under hybrid conditions; spoof-specific fine-tuning improves separability but struggles with unseen compositions; and dataset-specific adaptation on HSAD yields large performance gains (AST $>$97\% accuracy, $\approx$99\% F1), though residual errors persist for complex hybrids. These findings demonstrate that fine-tuning alone is not sufficient—robust hybrid-aware benchmarks like HSAD are essential to expose calibration failures, model biases, and factors affecting spoof detection in adversarial environments. HSAD thus provides both a dataset and an analytic framework for building resilient and trustworthy voice authentication systems.  
\end{abstract}


%% file: 01-introduction.tex
\begin{figure}
    \centering
    \includegraphics[width=1\linewidth]{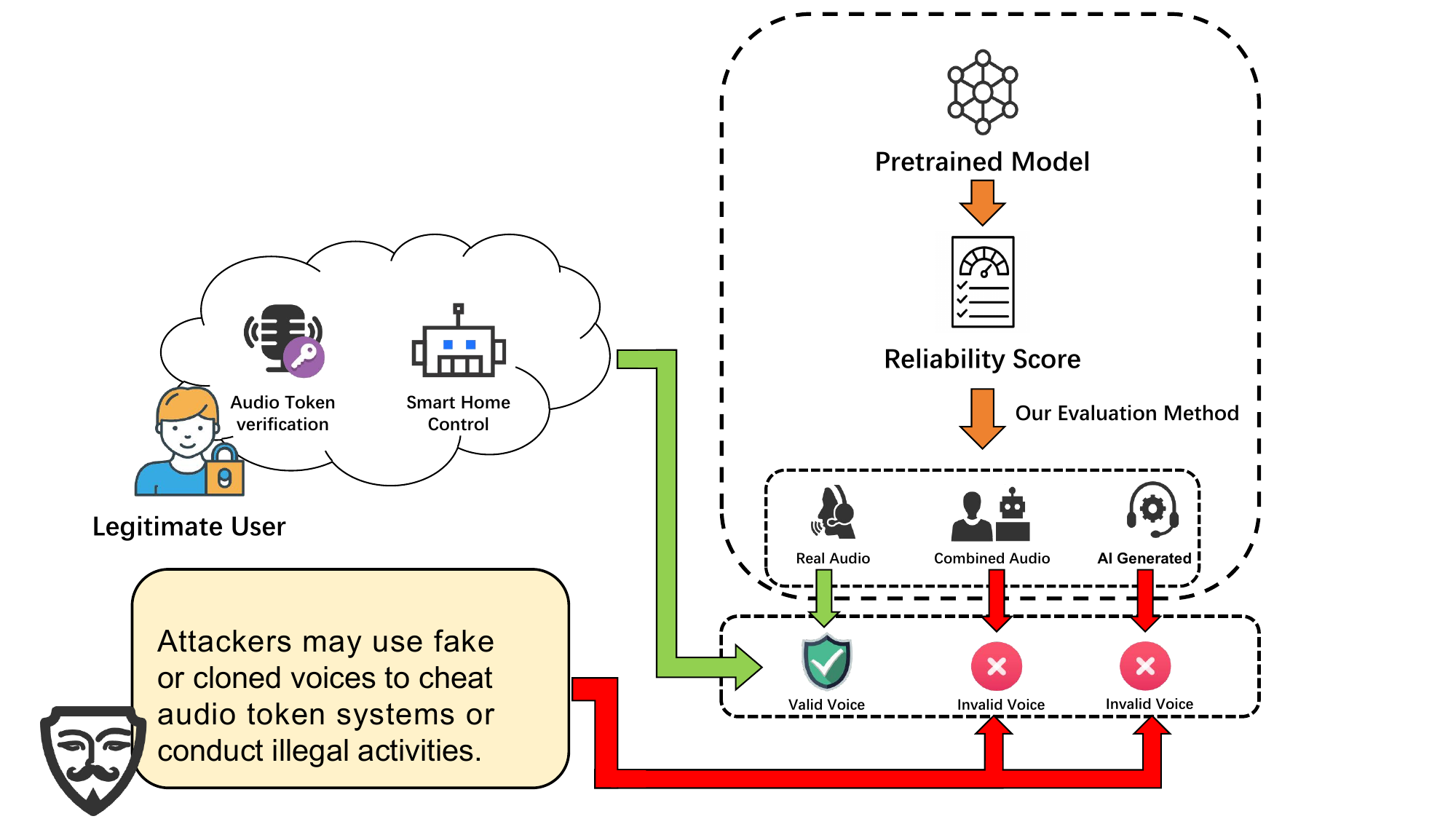}
    \caption{Our evaluation framework in a smart home audio token system. It highlights the challenge of detecting AI-synthesized, cloned, and hybrid speech that bypasses fine-tuned models, ensuring resilient control in adversarial environments.}
    \label{fig:audio-token-system}
\end{figure}

\vspace{-0.3cm}
\section{Introduction}

Voice authentication has rapidly become a cornerstone of secure access in smart homes, banking, and mobile applications. Its convenience, low cost, and contactless nature make it an attractive alternative to passwords or biometrics such as fingerprints~\cite{Ibrar2023}. At the same time, speech-driven interfaces are now embedded in everyday environments, from virtual assistants to telemedicine and financial services~\cite{Kim2016}.  

Yet, these systems are increasingly under attack. Advances in AI-driven voice cloning and zero-shot text-to-speech (TTS) now enable adversaries to synthesize speech that is nearly indistinguishable from real voices~\cite{Liu2023}. In high-profile cases, attackers have exploited cloned voices to trick financial institutions, leading to multi-million-dollar losses~\cite{Brewster2021}. Traditional countermeasures—often designed around binary genuine/spoof classification—are ill-prepared to handle the complexity of these evolving threats, particularly when genuine and synthetic speech are mixed within a single utterance.

Research in automatic speaker verification (ASV) has explored handcrafted features~\cite{Todisco2017cqcc}, CNN/DNN-based models~\cite{Zhang2021cnnspoof, Alam2021asvspoof}, and transformer architectures such as AST~\cite{gong2021astaudiospectrogramtransformer} and self-supervised encoders like Wav2Vec2 and HuBERT~\cite{baevski2020wav2vec, hsu2021hubert}. While these approaches show strong performance on clean benchmarks, they often fail when exposed to hybrid or adversarial spoofing conditions. A recurring assumption is that fine-tuning on an existing dataset is sufficient for generalization. However, as we show, this assumption does not hold in practice: models can remain overconfident yet brittle, misclassifying genuine speech or collapsing on unseen hybrids even after extensive adaptation.

One key bottleneck lies in the datasets themselves. Widely used corpora such as ASVspoof 2019 and 2021~\cite{nautsch2021asvspoof} are limited to binary or ternary spoof categories, and lack the hybrid compositions or realistic degradations present in deployment. While HAD~\cite{yi2021half} and ADD2023-PF~\cite{yi2023add} begin addressing segment-level manipulations, they remain narrow in scope and do not capture the diversity of real-world spoofing attacks. As a result, fine-tuned models evaluated on these datasets provide an incomplete picture of robustness.

To address this gap, we introduce the \textbf{Hybrid Spoofed Audio Dataset (HSAD)}, explicitly designed to stress-test detection under hybrid and adversarial conditions. HSAD contains 1,248 clean and 1,248 degraded utterances across four classes—human, cloned, zero-shot AI-generated, and hybrid compositions—annotated with spoof type, segment boundaries, and degradation metadata. Unlike prior corpora, HSAD introduces complex mixtures of real and synthetic speech alongside codec compression, channel filtering, and noise injection, better reflecting deployment environments.  

We benchmark five transformer-based models—including AST variants (MIT-AST, MattyB95-AST) and self-supervised encoders (Wav2Vec2, HuBERT)—to ask a critical question: \emph{when does fine-tuning fail, and what does it take to achieve robust spoof detection?} Our results show that while fine-tuning on HSAD improves performance dramatically ($>97\%$ accuracy, $\approx99\%$ F1), failures persist in hybrid and degraded settings, revealing calibration issues invisible on binary clean datasets. These lessons highlight the need for hybrid-aware datasets, richer annotations, and evaluation beyond raw accuracy.

The contributions of this paper are:
\begin{itemize}

    \item We construct the HSAD, a benchmark of 1,248 clean and 1,248 degraded utterances across four classes (human, cloned, AI-generated, hybrid), enriched with segment-level and degradation metadata.  

    \item  Through systematic evaluation of six transformer-based models (e.g., AST, Wav2Vec2, HuBERT), we show that pretrained models overgeneralize, while fine-tuned baselines still fail to generalize to hybrid or unseen attacks.  

    \item We identify key factors impacting detection, such as class overlap, segment mixing, and noise robustness, that reveal why fine-tuning alone is insufficient for reliable spoof detection.  

    \item  Our findings suggest that future systems must integrate hybrid-aware datasets, spoof-specific calibration, and multi-level feature modeling to achieve resilience in real-world deployments.  
\end{itemize}

%% file: 02-Relatedwork.tex
\vspace{-0.3cm}
\section{Related Work}

\subsection{Deepfake Audio Detection Models}

Early efforts in deepfake and spoofed speech detection primarily employed traditional signal processing and statistical approaches such as Gaussian Mixture Models (GMM)~\cite{wen2022multi}. While these methods showed initial success on datasets like ASVspoof 2015 and 2019, they struggled to model long-range dependencies and complex spoofing patterns, particularly under replay-based or hybrid attacks. Their reliance on fixed acoustic features (e.g., MFCCs, CQCCs) limited adaptability to emerging spoofing methods and real-world variability.

With the advent of deep learning, more expressive architectures were proposed. Light CNN (LCNN)~\cite{nautsch2021asvspoof} and LSTM-based~\cite{cheng2023analysis} classifiers introduced end-to-end learning frameworks that improved the modeling of temporal and local spectral structures. These models achieved stronger baselines across logical and physical access scenarios in ASVspoof challenges. However, their convolutional or recurrent inductive biases limited their ability to generalize across audio domains or handle composite utterances with mixed spoofing artifacts.

A significant leap forward occurred with the introduction of the Audio Spectrogram Transformer (AST)~\cite{gong2021astaudiospectrogramtransformer}. AST, modeled after the Vision Transformer (ViT), interprets audio spectrograms as visual patches and applies self-attention to capture global time-frequency relationships. AST achieved state-of-the-art performance on standard audio classification benchmarks such as ESC-50 (95.6\%) and Speech Commands V2 (97.4\%), proving its strength in generalization and long-range modeling. Importantly, AST's patch-based design facilitates interpretability and flexible handling of variable-length inputs, making it suitable for spoof detection.

Building upon AST, the SSAST-CL model~\cite{goel2024towards} incorporated contrastive pretraining to enhance representations learned from large-scale unlabeled audio. On the ASVspoof 2021 Logical Access dataset, SSAST-CL achieved an Equal Error Rate (EER) of 4.74\%, surpassing many supervised CNN-based baselines and demonstrating the efficacy of self-supervised contrastive learning for spoofing tasks.

Parallel work has explored joint modeling of Automatic Speaker Verification (ASV) and anti-spoofing Countermeasures (CM). Kanervisto et al.~\cite{kanervisto2022optimizing} proposed a reinforcement learning-based optimization strategy targeting the tandem Detection Cost Function (t-DCF), achieving a 20\% reduction in t-DCF relative to sequentially trained ASV+CM systems. Their method improved robustness on challenging ASVspoof 2019 logical access attacks, especially the low-confidence attack categories (A17--A19).

Recent approaches have also incorporated physical-layer acoustic modeling to address real-world variability. Micro-signature modeling of microphone hardware imperfections~\cite{patil2022microsignatures}, and spectral-temporal modulation analysis~\cite{cheng2023analysis}, have improved performance under replay and multi-stage tampering scenarios. These methods emphasize fine-grained signal anomalies often missed by conventional classifiers.

Furthermore, self-supervised transformer-based models such as Wav2Vec2 and HuBERT~\cite{baevski2020wav2vec,hsu2021hubert} have been increasingly adopted for spoof detection due to their strong speech representation learning capabilities. Although initially trained for ASR, recent work has shown that with modest task-specific fine-tuning, these models can rival specialized detection models in identifying synthetic and replayed speech. Their use of contrastive or masked prediction objectives enables effective pretraining on large-scale unlabeled corpora, providing a powerful foundation for downstream spoof detection tasks.

Despite these advances, current systems still face critical limitations. Most deepfake detectors assume binary spoof categories and fail to accommodate hybrid utterances composed of real and fake segments. Few models have been evaluated under spectrally degraded or temporally inconsistent conditions—yet these factors are pervasive in real-world telephony, conferencing, and streaming environments. Moreover, the interpretability of spoof classification remains underexplored, particularly for multi-stage and segment-level forgery detection.

\subsection{Spoofed Audio Datasets}

Progress in spoof detection research has largely depended on the availability of high-quality, diverse datasets. Early benchmarks such as the FoR-Original dataset~\cite{reimao2019for} focused on binary classification between bona fide and synthetic samples generated by conventional TTS systems. However, these datasets lacked variety in spoofing types and failed to simulate realistic deployment conditions such as background noise or hybrid manipulations.

The ASVspoof series (2015, 2019, 2021)~\cite{nautsch2021asvspoof} significantly advanced the field by introducing multiple spoofing modalities across logical (synthesis/voice conversion) and physical (replay) access channels. These corpora have served as gold standards for benchmarking spoofing countermeasures. Nevertheless, their utterance-level spoofing labels prevent them from modeling composite audios where only segments are manipulated. Moreover, their homogeneity in utterance structure and speaker identity limits generalization to spontaneous speech or multilingual environments.

To address partial forgery detection, the HAD dataset~\cite{yi2021half} introduced localized tampering by replacing specific words in human speech with synthetic ones. While useful for evaluating word-level detection, HAD's scope was restricted to short English utterances and TTS-based manipulations. ADD2023-PF~\cite{yi2023add} expanded upon this idea by providing segment-level annotations of tampered regions. However, it suffers from limited documentation on spoofing strategies, inconsistent spoof segment durations, and lack of replay or adversarial variants, making it challenging for robust evaluation across domains.

Other efforts have introduced adversarially perturbed speech datasets~\cite{wu2020adversarial}, and partially fake corpora that target signal-level spoofing rather than segment composition~\cite{alam2022partial}. Although these datasets improve upon the diversity of attack vectors, they often fail to simulate natural audio dynamics or include detailed annotations about the tampering process.

Moreover, existing datasets rarely contain hybrid utterances combining multiple spoofing techniques—such as cloned-replay, AI-generated-clone, or intra-speaker mixtures. Such hybrids are increasingly common in adversarial scenarios where attackers combine generative models with physical replay to defeat conventional detectors.

To address these shortcomings, the proposed Hybrid Spoofed Audio Dataset (HSAD) introduces a new level of granularity and realism. It features six hybrid spoof configurations, including intra-speaker mixed utterances, cloned-replay sequences, and AI-generated speech segments embedded in human recordings. Metadata includes speaker identity, gender, age group, spoofing type, segment start/end boundaries, codec application, and noise profile. Standardized 6-second durations across samples ensure compatibility with transformer-based models and downstream tasks requiring fixed-length input.

By bridging the gap between synthetic, real, and hybrid audio artifacts under real-world conditions, HSAD establishes a new benchmark for evaluating model robustness, generalization, and interpretability in spoofed speech detection.

%% file: 3-dataset-detection.tex
\vspace{-0.2cm}
\section{Dataset and Detection Methodologies}
\label{sec:dataset-detection}

\subsection{HSAD: Hybrid Spoofed Audio Dataset}

To evaluate spoof detection beyond binary setups, we introduce the Hybrid Spoofed Audio Dataset (HSAD). HSAD is designed to expose weaknesses of models that overfit to clean or narrow spoofing domains. It contains four audio categories: (1) genuine human speech, (2) cloned speech from Tacotron 2 with varying reference contexts, (3) fully AI-generated speech using zero-shot TTS, and (4) hybrid utterances mixing human and synthetic segments. Both clean and degraded versions are included to emulate real-world deployments.  

\textbf{Clean Human Speech:} We recorded 12 speakers (balanced by gender, aged 19–38) reading 104 sentences spanning eight linguistic categories (alphanumeric strings, coherent/incoherent pairs, error-embedded sentences, etc.). This ensured phonetic, semantic, and adversarial diversity.  

\textbf{Cloned and AI-generated Speech:} Each speaker’s voice was cloned under four reference conditions (C$_{1}$–C$_{4}$), from single-sentence embeddings to oracle-level target conditioning. Zero-shot synthesis was generated with multilingual, multi-speaker TTS prompts, matched semantically to human transcripts.  

\textbf{Hybrid Utterances:} To mimic tampering, we concatenated human and spoofed segments under three schemes: Human$\rightarrow$Synthetic, Synthetic$\rightarrow$Human, and Alternating. Cross-fade smoothing minimized transition artifacts.  

\textbf{Degradations:} To reflect practical conditions (telephony, conferencing), clean utterances were distorted with: (a) additive noise (10–30 dB SNR), (b) 4 kHz low-pass filtering, and (c) Opus compression (16/24 kbps).  

The resulting dataset includes 1,248 clean and 1,248 degraded utterances, evenly split across four classes. Each sample is annotated with speaker ID, spoof type, hybrid configuration, and degradation metadata.

\subsection{Detection Models}

We benchmark five transformer-based models adapted for four-class spoof detection (\{Human, Cloned, AI-generated, Hybrid\}).

\textbf{Spectrogram-based AST Models:}  
\begin{itemize}
    \item \textbf{AST-MIT}~\cite{gong2021astaudiospectrogramtransformer}: pretrained on AudioSet for broad acoustic coverage.  
    \item \textbf{AST-MattyB95}: fine-tuned on ASVspoof2019 LA, specialized for spoof detection.  
\end{itemize}
Both treat 128-bin log-Mel spectrograms (6s, 25 ms window, 10 ms hop) as image-like patches processed via a Vision Transformer encoder, with a 4-class softmax head. Transfer learning techniques include positional embedding resizing and classifier replacement.

\textbf{Waveform-based Self-Supervised Models:}  
\begin{itemize}
    \item Wav2Vec2-base-960h  
    \item Wav2Vec2-large-960h  
    \item HuBERT-base-ls960  
\end{itemize}
These models operate on raw waveform encodings, pretrained for ASR on LibriSpeech, and adapted with new 4-class heads. Training followed staged fine-tuning: initial layer freezing, progressive unfreezing, and optimization with AdamW and cosine decay.  

\subsection{Training and Evaluation Setup}

All models were trained on 6-second utterances standardized to 16 kHz, with speaker-disjoint 80/20 splits (10\% of training for validation). Optimizers were Adam/AdamW with cosine decay ($2 \times 10^{-5}$ init), batch size 3, for up to 20 epochs on an NVIDIA A100 GPU.  

We report \textbf{EER}, \textbf{min-tDCF}, accuracy, and F1-score. Confusion matrices reveal inter-class ambiguities, and class-wise breakdowns emphasize hybrid detection performance. For ASVspoof LA, we follow the official evaluation protocol; for HSAD, we analyze both clean and degraded subsets.  

\subsection{Design Advantages of HSAD}

Compared to prior corpora (e.g., ASVspoof2019~\cite{nautsch2021asvspoof}, HAD~\cite{yi2021half}), HSAD introduces:  
\begin{itemize}
    \item \textbf{Hybrid utterances:} exposing failures of fine-tuned models on mixed-content attacks.  
    \item \textbf{Controlled spoof fidelity:} varying clone conditions test sensitivity to embedding quality.  
    \item \textbf{Environmental robustness:} noise, codec, and channel artifacts simulate deployment.  
    \item \textbf{Rich metadata:} enabling explainable and interpretable spoof detection.  
\end{itemize}

This unified dataset+methods section directly ties HSAD’s design to the detection challenges and the observed phenomenon that fine-tuning alone is often insufficient under adversarial or hybrid conditions.

%% file: 03-Datasets.tex
\vspace{-0.2cm}
\section{Dataset Construction}
\label{sec:dataset-construction}

\subsection{Design Policy}

The proposed Hybrid Spoofed Audio Dataset (HSAD) is carefully constructed to benchmark the resilience and generalization capabilities of spoof detection systems under both clean and adversarial audio conditions. Recognizing the growing sophistication of speech forgery methods in real-world applications, the dataset incorporates four primary audio types: (1) genuine human speech, (2) cloned speech generated using speaker-conditional synthesis, (3) fully synthetic speech from zero-shot text-to-speech (TTS) systems, and (4) hybrid utterances mixing real and spoofed segments. In addition, noisy and degraded versions are produced to simulate real-world recording, transmission, and environmental distortions.

The dataset is split into two major versions to simulate diverse evaluation environments:

\begin{itemize}
    \item \textbf{Clean Hybrid Version:} Contains controlled recordings of genuine, cloned, synthetic, and hybrid audios, intended to isolate spoofing effects from environmental noise.
    \item \textbf{Noisy/Degraded Version:} Includes audio distorted via additive noise, codec compression, and spectral filtering, to reflect real-world challenges such as bandwidth limitations and microphone mismatch.
\end{itemize}

Each version is further divided into speaker-disjoint training, development, and test sets. To measure cross-attack generalization, the test set is subdivided into seen and unseen spoof types.

\subsection{Clean Real Audio Collection}

Human speech was recorded from 12 participants (balanced by gender; aged 19–38) in an acoustically treated environment using high-fidelity USB microphones. Recordings were captured at 44.1 kHz and downsampled to 16 kHz to align with standard audio spoof detection protocols. To ensure diversity in phonetic coverage, syntactic complexity, and semantic structure, each speaker read 104 sentences drawn from 8 carefully designed linguistic categories:

\begin{itemize}
    \item Alphanumeric Combination Sentences
    \item Pure Alphabetic Strings
    \item Numerical Sequences
    \item Standard English Phrases (e.g., New Concept English)
    \item Semantically Coherent Sentence Pairs
    \item Semantically Unrelated Sentence Pairs
    \item Sentences with Grammatical Errors
    \item Sentences with Semantic-Grammatical Anomalies
\end{itemize}

This content was chosen to emulate both natural and adversarial text styles, supporting robust spoof detection and interpretability studies.

\begin{table}[ht]
\centering
\resizebox{\linewidth}{!}{%
\begin{tabular}{@{}lr@{}}
\toprule
\textbf{Type Name} & \textbf{Number of Sentences} \\ 
\midrule
Alphanumeric Combination Sentences & 8 \\
Pure Alphabetic Sentences & 8 \\
Numeric Sequence Sentences & 8 \\
New Concept English Lesson Sentences & 16 \\
Semantically Coherent Sentence Pairs & 16 \\
Semantically Unrelated Sentences & 16 \\
Grammatical Error-Embedded Sentences & 16 \\
Semantic-Grammatical Anomaly Sentences & 16 \\
\bottomrule
\end{tabular}
}
\caption{Composition of sentence types in the human speech subset.}
\label{tab:model_performance}
\end{table}

\subsection{Cloned and Synthetic Audio Generation}

To simulate speaker-matched spoofing attacks, we employed a Tacotron 2-based TTS pipeline enhanced with speaker embeddings. Each participant's voice was cloned using four distinct reference embedding conditions, reflecting different levels of speaker-specific context:

\begin{enumerate}[label=C$_{\arabic*}$:]
    \item Single reference sentence (minimal embedding)
    \item 16-sentence subset (intermediate context)
    \item Full speaker corpus (complete context)
    \item Embedding from the target sentence (oracle-level cloning)
\end{enumerate}

Cloned outputs were evaluated using the ERes2NetV2 speaker verification model~\cite{chen2024eres2netv2}, with all samples exceeding a 0.70 reliability score, indicating high-fidelity synthesis suitable for anti-spoofing evaluation.

For zero-shot synthesis, we used a publicly available TTS model with multilingual and multi-speaker capabilities. These synthetic samples were generated using novel speaker prompts, and their transcripts were semantically matched to human recordings to control for content-based confounds.

\subsection{Hybrid Audio Construction}

Hybrid audios simulate real-world tampering where human and synthetic speech are interleaved. To construct these, we segmented and concatenated genuine and spoofed speech into a single utterance using three patterns:

\begin{itemize}
    \item Human-to-Synthetic (H$\rightarrow$S)
    \item Synthetic-to-Human (S$\rightarrow$H)
    \item Alternating Human-Synthetic-Human (H$\leftrightarrow$S$\leftrightarrow$H)
\end{itemize}

All transitions were smoothed using a 10 ms cross-fade window to minimize discontinuities. Each hybrid sample was annotated with segment-level metadata, including speaker ID, segment order, source type (genuine or spoofed), and gender.

\subsection{Noisy and Codec-Augmented Dataset}

To mimic realistic deployment conditions such as telephony, conferencing, or mobile app usage, we created degraded variants of the clean dataset. Three distortion strategies were applied:

\begin{itemize}
    \item \textbf{Additive Noise:} We introduced ambient noises such as street traffic, restaurant sounds, and synthetic white noise at 10, 15, 20, and 30 dB SNR using noise profiles from public sound databases.
    \item \textbf{Channel Filtering:} A low-pass filter with a 4 kHz cutoff was applied to emulate microphone channel mismatch and low-fidelity capture.
    \item \textbf{Codec Compression:} Audio samples were encoded and decoded using the Opus codec at 16 and 24 kbps, simulating VoIP transmission artifacts.
\end{itemize}

Distorted files retained their original spoofing label, and additional metadata was provided for each file including the degradation type, SNR level, and compression bitrate.

\subsection{Dataset Statistics}

The resulting HSAD corpus is composed of:

\begin{itemize}
    \item \textbf{Clean Hybrid Dataset:} 1,248 utterances evenly distributed across 4 classes (Human, Cloned, AI-generated, Hybrid), with 312 examples per class.
    \item \textbf{Noisy/Degraded Dataset:} 1,248 utterances generated by applying noise and compression to the clean set, preserving label and structure.
\end{itemize}

Every participant contributed equally to each class to ensure gender and linguistic balance. Segment ordering and composition patterns were randomized to prevent overfitting to structural cues.

\begin{figure}[H]
    \centering
    \includegraphics[width=1\linewidth]{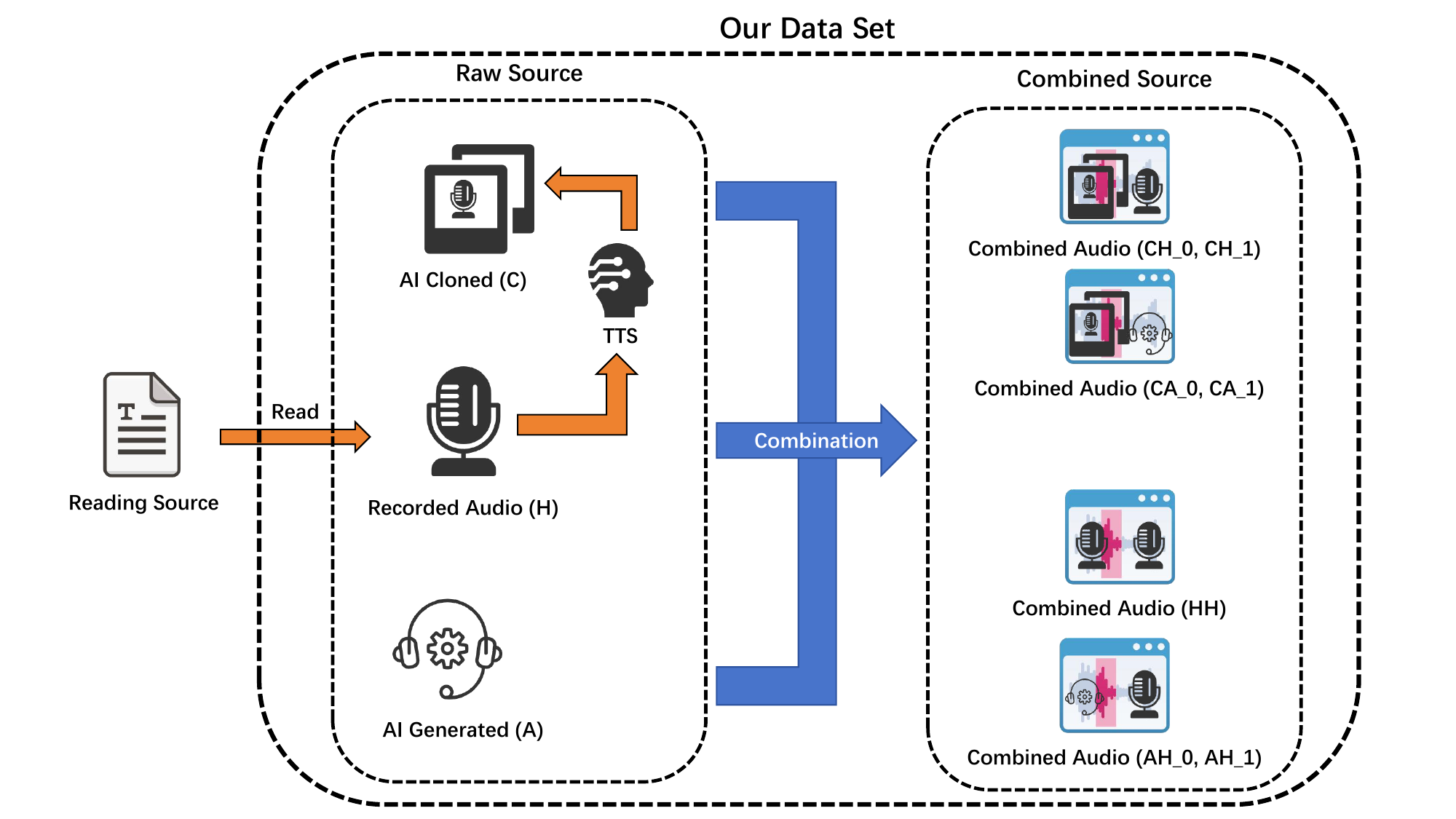}
    \caption{Construction pipeline of the Hybrid Spoofed Audio Dataset (HSAD), illustrating stages of genuine recording, cloning, synthesis, hybrid mixing, and noise augmentation.}
    \label{fig:dataset-framework}
\end{figure}
Figure~\ref{fig:dataset-framework} illustrates the full construction pipeline of the Hybrid Spoofed Audio Dataset (HSAD). It begins with the recording of clean human speech, followed by cloned speech synthesis using Tacotron 2 and zero-shot AI-based generation. Hybrid utterances are created by splicing genuine and spoofed segments in multiple arrangements. To simulate real-world conditions, the clean audio is augmented with noise, codec compression, and filtering. All generated samples are paired with metadata such as speaker identity, spoofing method, and degradation type, providing a comprehensive benchmark for evaluating spoof detection systems.

\subsection{Design Advantages}

Compared to prior corpora, the HSAD dataset introduces multiple novel aspects:

\begin{itemize}
    \item \textbf{Multi-source Hybrid Utterances:} Enables evaluation of detectors on utterances with both real and spoofed speech.
    \item \textbf{Fine-grained Spoof Fidelity Control:} Variable speaker reference conditions provide controlled difficulty levels for voice cloning.
    \item \textbf{Environmental Robustness Simulation:} Supports domain transfer testing with channel distortion and compression artifacts.
    \item \textbf{Rich Metadata Annotations:} Segment-wise labeling of source, gender, spoof type, degradation, and order enable explainable and interpretable model design.
\end{itemize}

These features fill critical gaps in existing benchmarks (e.g., ASVspoof 2019~\cite{nautsch2021asvspoof}, HAD~\cite{yi2021half}, and ADD2023-PF~\cite{yi2023add}), which either lack segment-level control, hybrid compositions, or realistic environmental conditions.

\subsection{Dataset Summary}

\begin{itemize}
    \item \textbf{Clean Hybrid Dataset:} 1,248 utterances (312 each of Human, Cloned, AI-generated, Hybrid), manually constructed from controlled human speech and cloned/synthetic models.
    \item \textbf{Noisy/Degraded Dataset:} 1,248 distorted utterances derived from the clean set with additive noise, low-pass filtering, and codec compression.
\end{itemize}

Each file is labeled with its spoofing class, speaker ID, gender, degradation settings (if applicable), and hybrid configuration metadata.

%% file: 04-Methodology.tex
\section{Detection Methodologies}

\subsection{Overview}

This study constructs a comprehensive hybrid spoofed audio dataset (HSAD) to support the detection of adversarial speech manipulations, including speech synthesis, voice conversion, and replay attacks~\cite{Ibrar2023}. To emulate real-world acoustic complexity, we developed six hybrid attack configurations, combining these modalities in sequences such as synthetic-replay and clone-replay. Cloned samples were synthesized using state-of-the-art TTS models, while replay segments were simulated via intra-speaker temporal concatenation, resulting in temporally fragmented and acoustically diverse signals.

To evaluate detection performance, we benchmarked five pretrained audio classification models grouped into two categories:

\begin{itemize}
    \item \textbf{Spectrogram-based Models:}
        \begin{itemize}
            \item \textbf{AST-MIT (MIT/ast-finetuned-audioset-10-10-0.4593)}~\cite{gong2021astaudiospectrogramtransformer}: A general-purpose AST model pretrained on AudioSet for broad acoustic coverage.
            \item \textbf{AST-MattyB95 (MattyB95/AST-ASVspoof2019)}: Fine-tuned on the ASVspoof2019 LA dataset for synthetic speech detection.
        \end{itemize}
    \item \textbf{Waveform-based Self-Supervised Models:}
        \begin{itemize}
            \item \textbf{Wav2Vec2-base-960h}~\cite{baevski2020wav2vec}
            \item \textbf{Wav2Vec2-large-960h}~\cite{baevski2020wav2vec}
            \item \textbf{HuBERT-base-ls960}~\cite{hsu2021hubert}
        \end{itemize}
\end{itemize}

All models were adapted with a 4-class softmax head to support fine-grained spoof classification into \textbf{Human}, \textbf{Cloned}, \textbf{AI-generated}, and \textbf{Hybrid} categories.

\subsection{Spectrogram-Based Architecture: Audio Spectrogram Transformer (AST)}

\subsubsection{Design and Architecture}

The AST model is a convolution-free architecture adapted from the Vision Transformer (ViT)~\cite{dosovitskiy2020vit} for audio spectrogram classification. It treats 128-bin log-Mel spectrograms as 2D images and segments them into overlapping $16 \times 16$ patches, projected into 768-dimensional embeddings. A positional encoding is added, and a [CLS] token is prepended.

The transformer encoder comprises 12 layers with 12 attention heads each. The final [CLS] token is classified into one of four spoofing categories. This architecture enables long-range modeling of time-frequency dependencies across entire utterances.

\begin{figure}[H]
    \centering
    \includegraphics[width=1\linewidth]{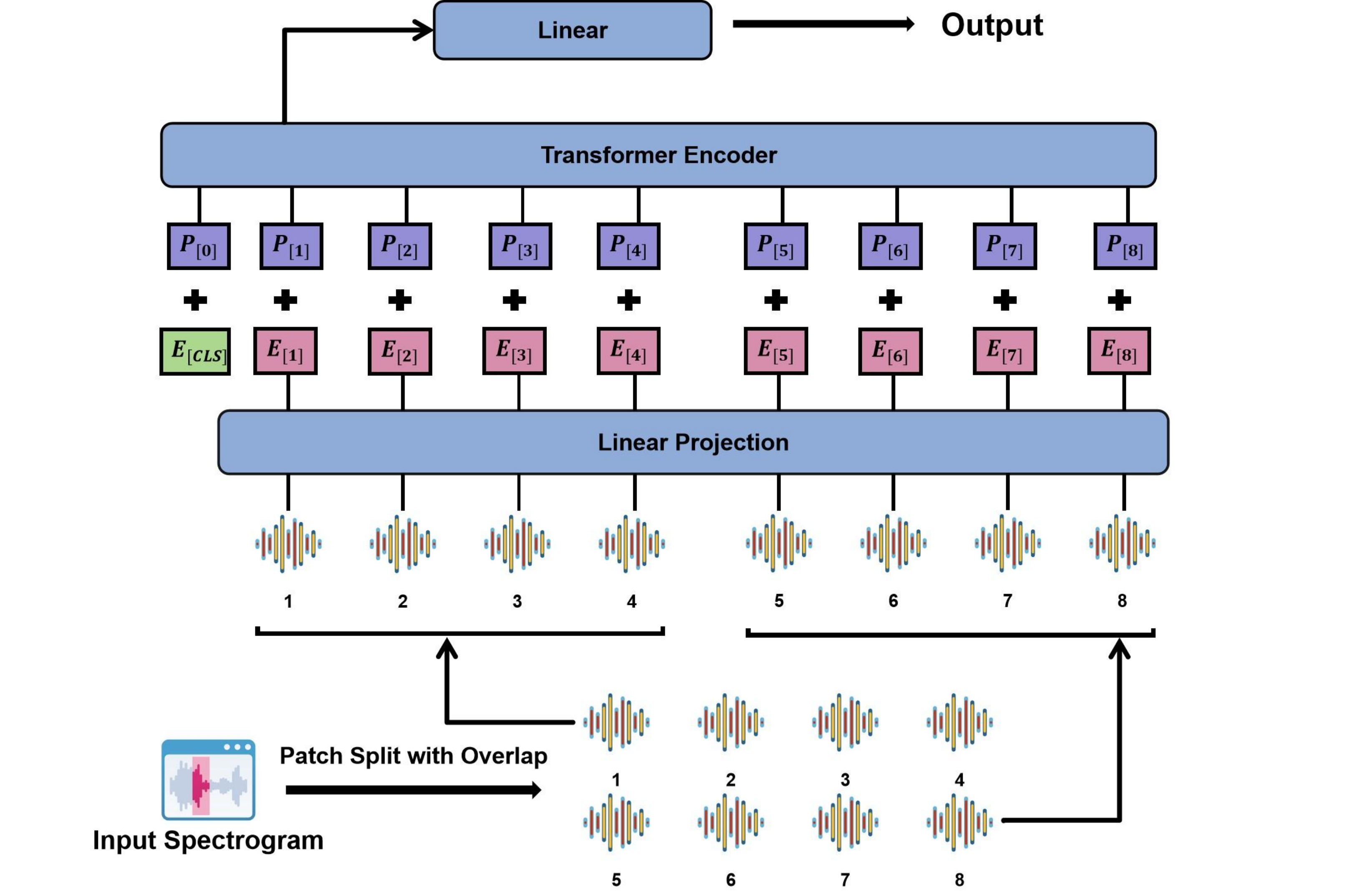}
    \caption{Architectural framework of the Audio Spectrogram Transformer (AST) model.}
    \label{fig:ast-architecture}
\end{figure}

\subsubsection{Pretraining and Transfer Learning Strategy}

We initialized AST-MIT using weights pretrained on AudioSet and AST-MattyB95 using weights from ASVspoof2019. To further adapt the models to our hybrid dataset, we applied the following ViT transfer learning techniques:

\begin{itemize}
    \item \textbf{Channel Adaptation:} ViT weights for RGB input were averaged to simulate single-channel spectrogram input.
    \item \textbf{Input Normalization:} Spectrograms normalized to zero mean, standard deviation 0.5.
    \item \textbf{Positional Embedding Resizing:} Bilinear interpolation reshaped 2D embeddings to spectrogram grid sizes.
    \item \textbf{Classification Head Replacement:} The ViT classifier was replaced with a 4-class spoof detection head.
\end{itemize}

We specifically used DeiT~\cite{touvron2021training} as the backbone, with 87M parameters pretrained on ImageNet via knowledge distillation.

\subsubsection{Training Configuration}

\begin{itemize}
    \item Optimizer: Adam with weight decay
    \item Learning rate: $2 \times 10^{-5}$ (cosine decay)
    \item Batch size: 3
    \item Epochs: 20 with early stopping
    \item Input: Zero-padded 6s log-Mel spectrograms (128 bins, 25ms window, 10ms hop)
\end{itemize}

\subsubsection{Label Definitions}

\begin{itemize}
    \item \textbf{Class 0 – Human:} Real speech recordings
    \item \textbf{Class 1 – Cloned:} TTS speech mimicking speaker identity
    \item \textbf{Class 2 – AI-generated:} Synthetic speech without identity control
    \item \textbf{Class 3 – Hybrid:} Concatenated human and synthetic segments
\end{itemize}

\subsection{Waveform-Based Architectures: Wav2Vec2 and HuBERT}

\subsubsection{Model Overview}

Three transformer-based speech encoders pretrained on waveform inputs were evaluated:

\begin{itemize}
    \item \textbf{Wav2Vec2-base-960h:} 7-layer CNN + 12 transformer layers, 8 heads, 768-dim~\cite{baevski2020wav2vec}.
    \item \textbf{Wav2Vec2-large-960h:} 24 transformer layers, 16 heads, 1024-dim (317M parameters).
    \item \textbf{HuBERT-base-ls960:} Architecturally similar to Wav2Vec2-base but trained to predict clustered speech units~\cite{hsu2021hubert}.
\end{itemize}

\begin{figure}[H]
    \centering
    \includegraphics[width=1\linewidth]{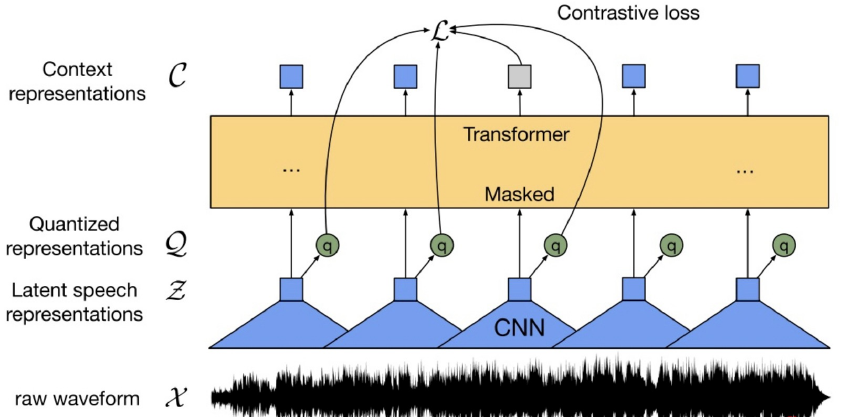}
    \caption{Architectural framework of the Self-Supervised Speech Models baseline.}
    \label{fig:model-def}
\end{figure}

\subsubsection{Transfer Learning and Fine-Tuning Strategy}

All waveform-based models were initialized with pretrained weights and adapted for spoof classification as follows:

\begin{itemize}
    \item A new 4-class classification head was appended.
    \item Layers were frozen for 5 epochs and then progressively unfrozen.
    \item Audio was padded to 6s, converted to log-Mel spectrograms (128 bins).
    \item Cross-entropy loss was minimized using AdamW~\cite{loshchilov2017decoupled}, with learning rate $2 \times 10^{-5}$, cosine annealing~\cite{loshchilov2016sgdr}, batch size 3.
\end{itemize}

\subsubsection{Label Space (Unified Across All Models)}

\begin{itemize}
    \item \textbf{0 – Human:} Authentic real-world speech
    \item \textbf{1 – Cloned:} Speaker-matched synthetic speech
    \item \textbf{2 – AI-generated:} Identity-agnostic synthetic speech
    \item \textbf{3 – Hybrid:} Mixed real and synthetic audio segments
\end{itemize}

%% file: 05-Experiment.tex
\vspace{-0.2cm}
\section{Evaluation}

\subsection{Experimental Setup}

The evaluation was conducted using both clean and noisy versions of the proposed hybrid spoofing dataset. Four spoofing classes were defined: (0) genuine human speech, (1) AI-cloned speech, (2) AI-generated speech, and (3) hybrid (mixed-source) speech. These classes reflect realistic adversarial scenarios designed to challenge anti-spoofing models under varied linguistic, acoustic, and synthesis conditions.

Audio recordings were sampled at 16 kHz and converted into 128-bin log-Mel spectrograms using a 25 ms Hamming window and a 10 ms frame shift. All experiments were conducted on an NVIDIA A100 GPU. Training followed an 80/20 train-test split, ensuring speaker disjointness to prevent overfitting.

Each model was trained for 20 epochs using the Adam optimizer with an initial learning rate of $2 \times 10^{-5}$, cosine decay scheduling, and early stopping. Batch size was fixed at 3 for stability given the variable input lengths and memory constraints.

\subsection{Evaluation Metrics}

Performance was measured using classification accuracy, F1-score, false positive rate (FPR), and false negative rate (FNR) across the four spoofing classes. For reliability-based binary classification (used in baseline benchmarking), the following thresholding rule was applied:

\[
\hat{y} =
\begin{cases}
0 & \text{if } |\text{real\_tag} - \text{reliability\_score}| < 0.5 \\
1 & \text{otherwise}
\end{cases}
\]

Overall classification accuracy was computed as:

\[
\text{Accuracy} = \frac{C}{N} \times 100\%
\]

where \(C\) is the number of correct predictions and \(N\) is the total number of samples. Confusion matrices and reliability score distributions were used to analyze error patterns and inter-class ambiguities.

\subsection{Datasets}

To establish robust benchmark performance for spoofed speech detection, we conducted evaluations on two datasets: the widely used ASVspoof 2019 Logical Access (LA) dataset~\cite{Vestman2019} and our newly constructed hybrid spoofed audio dataset, introduced in Section~\ref{sec:dataset-construction}.

\textbf{ASVspoof 2019 LA Dataset.} The ASVspoof 2019 LA dataset serves as a standard benchmark in the field of automatic speaker verification (ASV). It comprises bona fide and spoofed utterances generated using advanced text-to-speech (TTS) and voice conversion (VC) algorithms. The dataset is organized into disjoint training, development, and evaluation sets, offering a comprehensive framework to assess generalization under realistic synthesis conditions.

\textbf{Proposed Hybrid Spoofed Audio Dataset (HSAD).}  
To overcome the limitations of existing benchmarks—such as binary classification focus, limited spoof type diversity, and the absence of hybrid composition realism—we constructed the HSAD dataset. It includes six carefully designed categories that reflect a spectrum of real-world and adversarial audio conditions:

\begin{itemize}
    \item \textbf{G1 – Genuine Human:} Natural, untouched speech recordings from real speakers.
    \item \textbf{G2 – Pure AI Clone:} Cloned speech generated using Tacotron 2-based models with varying levels of speaker reference embeddings (e.g., single-sentence, corpus-level, and target-matched).
    \item \textbf{G3 – Pure AI Generated:} Fully synthetic speech created via zero-shot TTS systems with no prior speaker conditioning.
    \item \textbf{G4 – Mixed: AI Generated + Human:} Spliced utterances formed by concatenating segments of AI-generated and human speech to simulate content injection attacks.
    \item \textbf{G5 – Mixed: AI Cloned + AI Generated:} Utterances combining cloned and AI-generated segments, introducing complex spoofing configurations.
    \item \textbf{G6 – Human Recombined:} Human-only segments rearranged into hybrid-like flows to emulate natural conversational variation without introducing spoof artifacts.
\end{itemize}

Each audio sample is labeled with speaker ID, spoof type, segment structure, and signal fidelity level. Beyond clean speech, the dataset also includes an adversarial variant with environmental noise (10–30 dB SNR), channel filtering (low-pass at 4 kHz), and compression artifacts (Opus codec at 16–24 kbps) to simulate deployment conditions. These features support detailed robustness evaluation and promote generalization studies for both fine-grained classification and real-world ASV defense mechanisms.

\subsection{Baseline Models}

To evaluate performance on the ASVspoof 2019 LA and HSAD datasets, we selected six state-of-the-art transformer-based models from the Hugging Face model repository, spanning a spectrum of general-purpose, domain-specific, and self-supervised speech models.

\textbf{MIT-AST}~\cite{gong2021astaudiospectrogramtransformer}: A general-purpose Audio Spectrogram Transformer (AST) model pretrained on the AudioSet corpus using weakly labeled 10-second audio clips. This model captures a wide range of acoustic events and serves as a robust baseline for evaluating generalization across diverse sound conditions.

\textbf{MattyB95}~\cite{MattyB95ASTASVspoof5}: A domain-specialized AST model fine-tuned on the ASVspoof 2019 Logical Access dataset. Optimized for detecting synthetic speech, it provides a strong benchmark for binary spoofing classification tasks.

\textbf{WpythonW}~\cite{WpythonW}: This AST variant is trained exclusively on ElevenLabs-generated synthetic speech, offering insights into model generalization across spoofing techniques not covered in standard benchmarks such as ASVspoof.

\textbf{Wav2Vec2-base-960h}~\cite{baevski2020wav2vec}: A self-supervised model pretrained on 960 hours of Librispeech using contrastive learning objectives. This base variant has ~95M parameters and is commonly used for ASR and representation learning. It provides a baseline for assessing feature extractor robustness in spoof detection without task-specific tuning.

\textbf{Wav2Vec2-large-960h}~\cite{baevski2020wav2vec}: A larger variant of Wav2Vec2 (~317M parameters), also trained on Librispeech. Due to its larger capacity and deeper architecture, it enables evaluation of scalability and expressive power in speech classification tasks.

\textbf{HuBERT-base-ls960}~\cite{hsu2021hubert}: The HuBERT (Hidden-Unit BERT) base model trained on 960 hours of Librispeech via masked prediction of clustered acoustic units. HuBERT combines self-supervised pretraining with improved representation learning over Wav2Vec2 and provides a valuable comparison point for hybrid spoof classification.

Together, these models encompass a range of architectural and training paradigms:
\begin{itemize}
    \item AST models: Focus on time-frequency self-attention from spectrograms.
    \item Wav2Vec2 and HuBERT: Operate directly on raw waveforms via self-supervised learning.
    \item Domain-specific vs. general-purpose: Comparison across generalization and spoofing specialization.
\end{itemize}

This comprehensive baseline selection allows us to assess model robustness, generalization, and discriminative performance under both standard spoofing (ASVspoof) and complex hybrid adversarial conditions (HSAD).

\subsection{Performance on ASVspoof 2019 LA}

To evaluate baseline spoof detection performance, we tested three transformer-based models on the ASVspoof 2019 Logical Access (LA) dataset~\cite{Vestman2019}. These included MIT-AST, a general-purpose Audio Spectrogram Transformer (AST) pretrained on AudioSet; MattyB95, specifically fine-tuned on ASVspoof 2019 challenge data; and WpythonW, trained on ElevenLabs synthetic speech. Table~\ref{tab:model_performance} presents the number of correct predictions and accuracy for each model.

\begin{table}[t]
    \caption{Performance on ASVspoof 2019 LA Dataset.}
			\vspace{-0.2cm}
	\centering
	\small
    
	\setlength{\extrarowheight}{0.5pt}
	\begin{tabular}
		{m{1.7cm}<{\centering} m{2.2cm}<{\centering} m{1.9cm}<{\centering}} 
		\Xhline{1.5\arrayrulewidth}
		\textbf{Model Name} &\textbf{Correct Predictions} &\begin{tabular}{@{}c@{}}\textbf{Accuracy ( \%)}\end{tabular}\\
		\Xhline{1.5\arrayrulewidth}
        MIT-AST & 63,663 / 71,237 & 89.37\% \\
        MattyB95 & 63,863 / 71,237 & 89.65\% \\
        WpythonW & 42,143 / 71,237 & 59.16\% \\
		\Xhline{1.5\arrayrulewidth}
	\end{tabular}
		\vspace{-0.4cm}
	\label{table:datasets}
\end{table}
\vspace{-0.1cm}

The results show that transformer-based architectures are highly effective for speech spoof detection. Both MIT-AST and MattyB95 achieved high accuracy, with MattyB95 slightly outperforming MIT-AST due to domain-specific fine-tuning. This highlights the advantage of combining large-scale pretraining with task-specific adaptation.

In contrast, the WpythonW model—despite its similar architecture—showed markedly lower performance. Trained only on ElevenLabs-generated samples, its poor generalization underscores the danger of narrow domain bias in anti-spoofing systems. These findings emphasize the necessity of diverse training data and hybrid-aware fine-tuning when targeting real-world spoofing threats.


\subsection{Performance on the Proposed HSAD Dataset}

\subsubsection{\textbf{MIT-AST and MattyB95 Models}}
\hfill\\
\textbf{1) Pretrained models}

To assess the generalization capabilities of transformer-based models under complex spoofing conditions, we evaluated MIT-AST and MattyB95 on the HSAD dataset. Table~\ref{table:datasets-1} shows reliability score statistics across the six spoofing groups.

Table~\ref{table:datasets-1} summarizes the reliability score statistics for both models across six spoofing groups: \textbf{G1} (Genuine Human), \textbf{G2} (Pure AI Clone), \textbf{G3} (Pure AI Generated), \textbf{G4} (Mixed: AI Generated + Human), \textbf{G5} (Mixed: AI Clone + AI Generated), and \textbf{G6} (Human Recombined).

\begin{table}[t]
\caption{Reliability score statistics for six spoofing groups (G1–G6) evaluated using MIT-AST and MattyB95 models on the proposed HSAD dataset.}
\vspace{-0.3cm}
\centering
\small
\begin{tabular}{m{0.2cm}<{\centering} m{0.7cm}<{\centering} m{1.0cm}<{\centering} m{0.7cm}<{\centering} m{0.7cm}<{\centering} m{0.9cm}<{\centering} m{0.9cm}<{\centering}}
\Xhline{1.5\arrayrulewidth}
\quad & \textbf{Group} & \textbf{Mean} & \textbf{Std Dev} & \textbf{Max} & \textbf{Min} & \textbf{Mode} \\
\Xhline{1.5\arrayrulewidth}
\multirow{6}*{\rotatebox{90}{MIT-AST}} 
& G1 & 0.8414 & 0.1339 & 0.9768 & 0.0071 & 0.7977 \\ 
& G2 & 0.8756 & 0.0829 & 0.9923 & 0.0185 & 0.8828 \\ 
& G3 & 0.8606 & 0.0959 & 0.9639 & 0.3924 & 0.7320 \\ 
& G4 & 0.8009 & 0.1398 & 0.9827 & 0.0908 & 0.7479 \\ 
& G5 & 0.8092 & 0.1343 & 0.9927 & 0.0185 & 0.7705 \\
& G6 & 0.7904 & 0.1748 & 0.9794 & 0.0095 & 0.0095 \\ 
\Xhline{1.0\arrayrulewidth}
\multirow{6}*{\rotatebox{90}{MattyB95}} 
& G1 & 0.5385 & 0.4925 & 1.0000 & 5.96e-7 & 5.96e-7 \\
& G2 & 0.9821 & 0.1319 & 1.0000 & 5.96e-7 & 1.0000 \\ 
& G3 & 0.3519 & 0.3519 & 1.0000 & 5.96e-7 & 5.96e-7 \\ 
& G4 & 0.3669 & 0.4733 & 1.0000 & 5.96e-7 & 5.96e-7 \\ 
& G5 & 0.6658 & 0.4679 & 1.0000 & 5.96e-7 & 5.96e-7 \\ 
& G6 & 0.2713 & 0.4435 & 1.0000 & 5.96e-7 & 5.96e-7 \\
\Xhline{1.5\arrayrulewidth}
\end{tabular}
\label{table:datasets-1}
\vspace{-0.4cm}
\end{table}

The MIT-AST model, despite achieving a high average accuracy of 93.67\% on the HSAD dataset, failed to correctly identify any genuine human utterances, misclassifying them as spoofed. This is evidenced by the close overlap of mean reliability scores for Group G1 (Human, 0.8414), G2 (Cloned, 0.8756), and G3 (Generated, 0.8606). Such overlap indicates insufficient class separation and undermines model interpretability and trustworthiness in real-world deployment.

In contrast, the MattyB95 model exhibited a lower overall accuracy of 65\% but showed improved class distinction. Its reliability scores for genuine human speech (G1: 0.5385) and human recombined segments (G6: 0.2713) were substantially lower than those for cloned (G2: 0.9821) and hybrid compositions, offering better separation between genuine and synthetic content.

However, both models struggled with the hybrid categories. Groups G4 (AI Generated + Human) and G5 (AI Clone + AI Generated) yielded highly dispersed scores with large standard deviations, suggesting confusion due to complex boundary conditions and mixed-source signal characteristics.

These results underscore three critical insights:

1) Limitations of Standard Pretraining: Models like MIT-AST, despite broad pretraining on AudioSet, are not calibrated for fine-grained spoof discrimination and tend to overgeneralize, especially in the presence of hybrid or partially spoofed audio.

2) Benefit of Spoof-Specific Tuning: While MattyB95 exhibits better discrimination between real and spoofed speech, its performance degrades under distribution shifts, such as unseen hybrid constructs.

3) HSAD Dataset Utility: The HSAD dataset introduces nuanced scenarios and spoofing combinations absent from traditional corpora, highlighting its importance in benchmarking robust, future-ready detection architectures.

In conclusion, existing models fail to reliably detect and classify hybrid spoofing attacks due to overlapping decision boundaries. These findings motivate the development of hybrid-aware models with finer temporal segmentation, semantic consistency modeling, and adaptive spoof calibration strategies tailored to composite real-world conditions.

\textbf{2. Fine-Tuned Models}

To validate the efficacy of the proposed HSAD dataset, we fine-tuned two transformer-based models—Model A, adapted from the MIT-AST architecture, and Model B, derived from MattyB95. Both models retained the Audio Spectrogram Transformer (AST) backbone and were retrained on our multi-source dataset to enhance domain-specific spoof detection.

Training was conducted for 20 epochs using the Adam optimizer and cosine learning rate scheduling, with early stopping based on validation loss. Audio inputs were standardized to 128-bin log-Mel spectrograms computed with 25ms frame length and 10ms hop size. A batch size of 3 was used, and training was performed on an NVIDIA A100 GPU.

Model A achieved a validation accuracy of 97.88\%, while Model B slightly exceeded this with 98.08\%. On the test set, both models stabilized at 97\% accuracy, substantially outperforming the baseline AST variants. The confusion matrices illustrate these improvements. Model A (fine-tuned MIT-AST) achieved particularly consistent performance across all categories, while Model B (fine-tuned MattyB95) showed strong discriminability among hybrid and synthetic speech types.

\begin{figure}[H]
    \centering
    \includegraphics[width=0.5\linewidth]{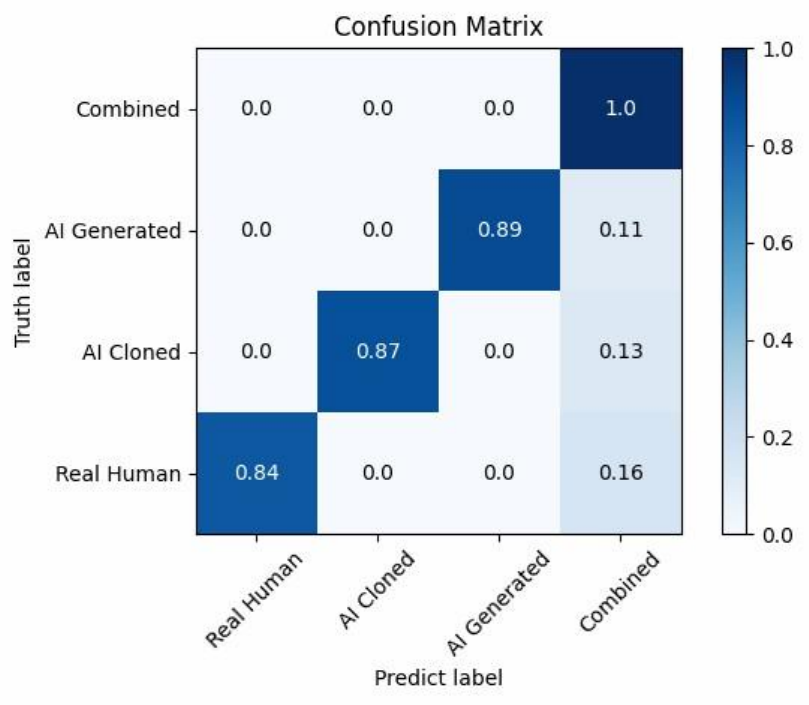}
    \caption{Confusion matrix – MIT fine-tuned (Model A)}
\end{figure}
\begin{figure}[H]
    \centering
    \includegraphics[width=0.5\linewidth]{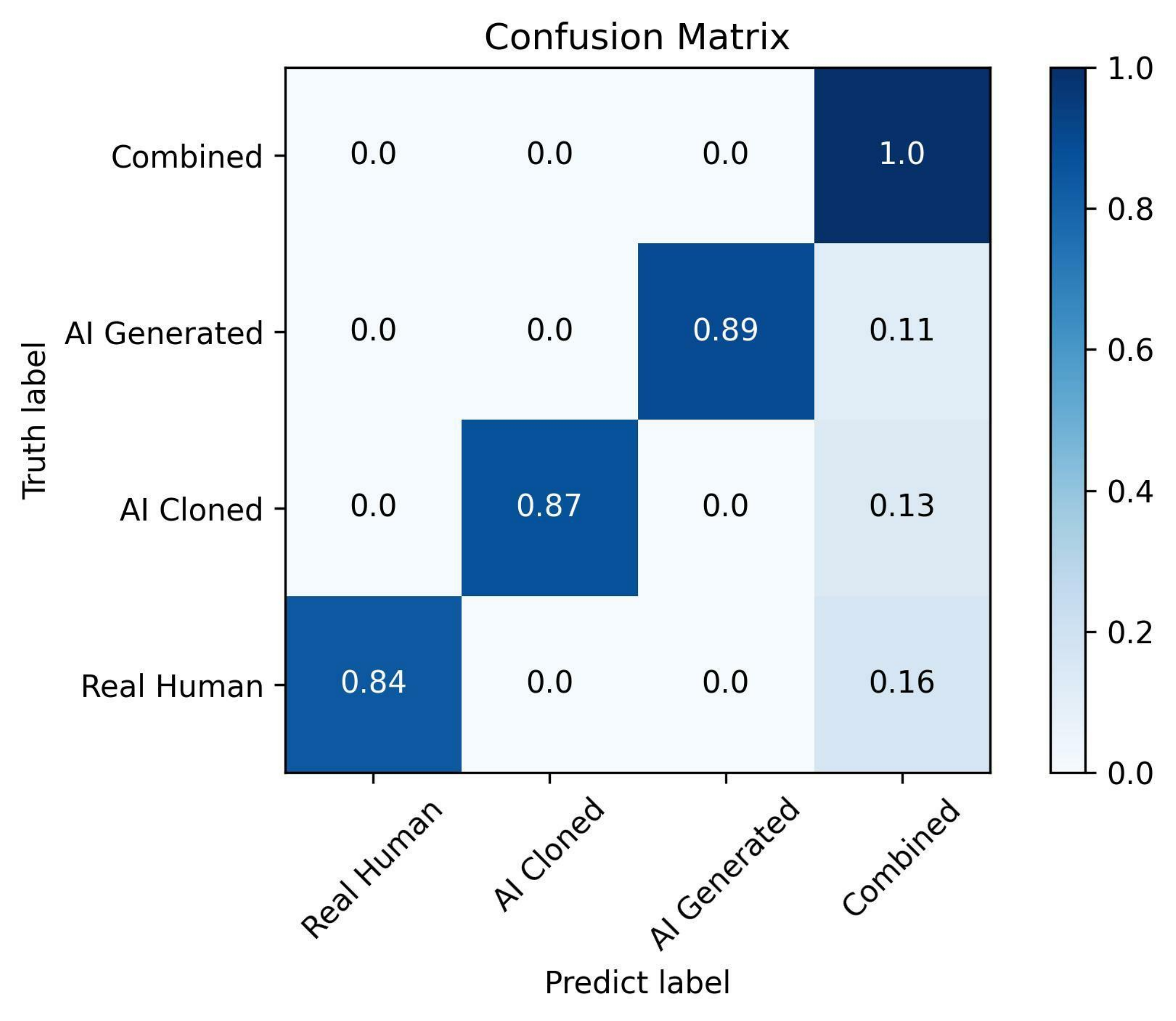}
    \caption{Confusion matrix – MattyB95 fine-tuned (Model B)}
\end{figure}

Both models correctly identified hybrid audio samples with 100\% precision, demonstrating their ability to capture compositional cues that standard classifiers often miss. For AI-generated and AI-cloned content, recall exceeded 88\%, confirming their capability to recognize nuanced synthetic speech even when integrated into natural audio flows. The classification accuracy for human speech rose to 84\%, representing a dramatic improvement compared to the baseline MIT-AST model, which had previously misclassified all genuine utterances.

Performance gains extended beyond accuracy. As shown in Fig.~\ref{fig:all_comb}, Model A achieved a 71\% reduction in false positives for human speech (from 1,207 to 197), while Model B demonstrated a 95\% reduction in false negatives for AI-generated content (from 14,117 to 684). Both models reached an F1-score of 99\%, indicating strong balance between precision and recall. Furthermore, Model A reduced its parameter count by 0.4M compared to the original MIT-AST (from 86.6M to 86.2M) while achieving a 2.27\% F1-score gain, highlighting the efficiency and effectiveness of dataset-specific adaptation.

\begin{figure}[t]
    \centering
    \includegraphics[width=\linewidth]{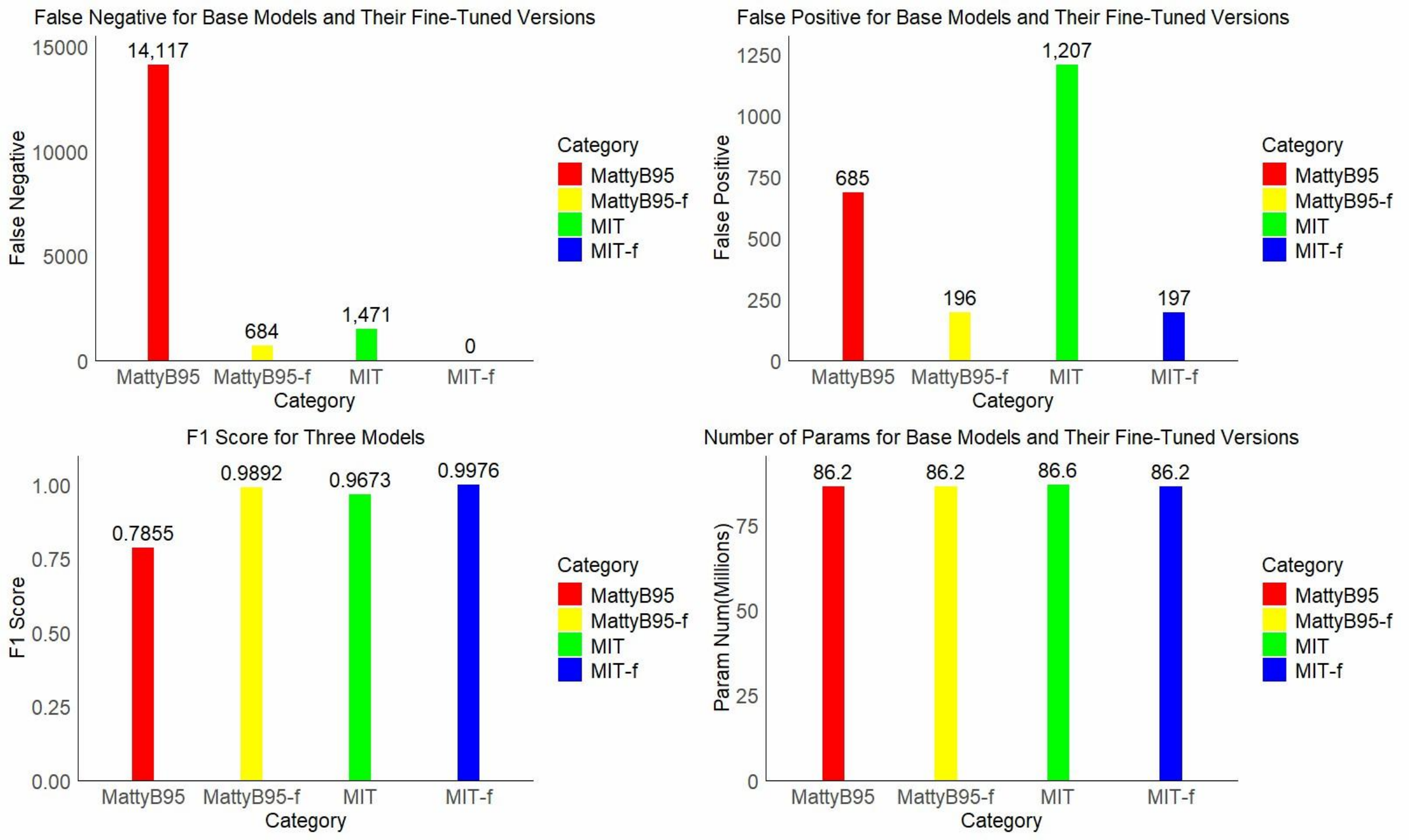}
    \caption{Comparison of false positives, false negatives, F1-score, and parameter count across baseline and fine-tuned models}
    \label{fig:all_comb}
\end{figure}

These findings emphasize that models pretrained on general-purpose audio datasets struggle with complex spoof compositions unless retrained on spoof-specific corpora. Although MIT-AST benefited from broad AudioSet pretraining, it lacked calibration for spoof detection tasks, particularly under hybrid scenarios. Conversely, the MattyB95 baseline model exhibited a stronger spoof detection bias but suffered from low reliability scores and misclassifications on unfamiliar or subtle combinations of synthetic and real audio. After fine-tuning, both models demonstrated significant improvements in generalization, particularly under adversarial and mixed-source conditions.

In conclusion, fine-tuning on the HSAD dataset not only improved classification accuracy and F1-score, but also enhanced the models’ robustness against false classifications and improved computational efficiency. These results affirm the importance of dataset-specific training for spoof detection and underline the potential of HSAD as a benchmark for advancing robust, real-world anti-spoofing solutions.

\begin{table}[t]
    \caption{Performance on the Proposed HSAD Dataset.}
			\vspace{-0.2cm}
	\centering
	\small
    
	\setlength{\extrarowheight}{0.5pt}
	\begin{tabular}
		{m{2.5cm}<{\centering} m{2.2cm}<{\centering} m{1.9cm}<{\centering}} 
		\Xhline{1.5\arrayrulewidth}
		\textbf{Model Name} &\textbf{Correct Predictions} &\begin{tabular}{@{}c@{}}\textbf{Accuracy ( \%)}\end{tabular}\\
		\Xhline{1.5\arrayrulewidth}
        Wav2vec2-base-960h & 1,697 / 8,549 & 20.06\% \\
        Wav2vec2-large-960h & 298 / 8,459 & 3.52\% \\
        Hubert-base-ls960 & 110 / 8,459 & 1.30\% \\
		\Xhline{1.5\arrayrulewidth}
	\end{tabular}
		\vspace{-0.4cm}
	\label{table:wave2_table1}
\end{table}
\vspace{-0.1cm}

\subsubsection{\textbf{Wav2vec2-base-960h, Wav2vec2-large-960h and Hubert-base-ls960 models}}
\hfill\\
\textbf{1. Pretranied Models}

To further examine the generalization capability of self-supervised speech encoders in spoof detection, we evaluated three widely used transformer-based models—Wav2Vec2-base-960h, Wav2Vec2-large-960h, and HuBERT-base-ls960—on the proposed Hybrid Spoofed Audio Detection (HSAD) dataset.

As shown in Table~\ref{table:wave2_table1}, these models exhibit poor classification performance in their pretrained form. Wav2Vec2-base-960h achieved the highest accuracy among the three, at only 20.06\%, while Wav2Vec2-large-960h and HuBERT-base-ls960 achieved only 3.52\% and 1.30\%, respectively. These results clearly demonstrate that general-purpose self-supervised pretraining—although effective for automatic speech recognition (ASR)—does not transfer well to the nuanced task of spoofed and hybrid audio detection without further adaptation.

The class-wise prediction distributions in Table~\ref{table:wave2_table2} reveal pronounced mode collapse and severe misclassification. Wav2Vec2-large-960h predicted nearly all test samples as G3 (AI-Generated), accounting for 97.67\% of its predictions. This extreme prediction bias suggests that the model latches onto dominant acoustic patterns present in synthetic speech and fails to generalize across diverse spoofing classes. Similarly, HuBERT-base-ls960 misclassified 51.16\% of inputs as G3 and 28.45\% as G1 (Human), indicating a confused semantic boundary between real and synthetic speech. Wav2Vec2-base-960h produced slightly more distributed predictions, but 53.56\% of its predictions still skewed toward G2 (AI-Cloned), while human and hybrid classes were largely ignored (e.g., G1: 0.43\%, G4: 1.61\%).

Table~\ref{table:wave2_table3} provides deeper insight into reliability score distributions across spoofing categories. All three models exhibit nearly flat mean scores ranging betwee
n 0.48 and 0.52 across G1–G6, with high standard deviations and widely varying maximum and minimum scores. This lack of class-specific confidence and high variance highlights the models’ inability to separate spoofed versus genuine segments or capture structural spoofing artifacts such as hybrid overlaps or recombinations. Their embeddings appear to be non-discriminative, yielding high-entropy, low-certainty decisions that are effectively equivalent to random guessing.

These findings confirm that pretrained Wav2Vec2 and HuBERT models, in their default form, are ill-equipped for hybrid spoof detection. Their representations lack sensitivity to subtle synthetic transitions and acoustic artifacts introduced by cloning, generative synthesis, or recombination. Without exposure to spoof-specific supervisory signals, these models develop biased, overly generic audio representations that fail to generalize to adversarial conditions. As such, effective spoof detection requires targeted fine-tuning on high-fidelity datasets like HSAD that explicitly model realistic, multi-source spoofing phenomena.

\begin{table}[t]
\caption{Reliability score statistics for six spoofing groups (G1–G6) evaluated using Wav2vec2-base-960h, Wav2vec2-large-960h and Hubert-base-ls960 models on the proposed HSAD dataset.}
\vspace{-0.3cm}
\centering
\small
\begin{tabular}{m{2.6cm}<{\centering} m{0.5cm}<{\centering} m{0.5cm}<{\centering} m{0.5cm}<{\centering} m{0.5cm}<{\centering} m{0.5cm}<{\centering} m{0.5cm}<{\centering}}
\Xhline{1.5\arrayrulewidth}
\textbf{Model Name} & \textbf{G1} & \textbf{G2} & \textbf{G3} & \textbf{G4} & \textbf{G5} & \textbf{G6} \\
\Xhline{1.5\arrayrulewidth}

Wav2vec2-base-960h& 0.43 & 53.56 & 0.00 & 1.61 & 18.96 & 11.25 \\ 
Wav2vec2-large-960h & 0.00 & 0.10 & 97.67 & 0.00 & 4.26 &0.00\\ 
Hubert-base-ls960& 28.45 & 0.00 & 51.16 & 2.21 & 0.00 & 0.00 \\ 
\Xhline{1.5\arrayrulewidth}
\end{tabular}
\label{table:wave2_table2}
\vspace{-0.4cm}
\end{table}

\begin{table}[t]
\caption{Reliability score statistics for six spoofing groups (G1–G6) evaluated using Wav2vec2-base-960h, Wav2vec2-large-960h and Hubert-base-ls960 models on the proposed HSAD dataset.}
\vspace{-0.3cm}
\centering
\small
\begin{tabular}{m{0.2cm}<{\centering} m{0.7cm}<{\centering} m{1.0cm}<{\centering} m{0.7cm}<{\centering} m{0.7cm}<{\centering} m{0.9cm}<{\centering} m{0.9cm}<{\centering}}
\Xhline{1.5\arrayrulewidth}
\quad & \textbf{Group} & \textbf{Mean} & \textbf{Std Dev} & \textbf{Max} & \textbf{Min} & \textbf{Mode} \\
\Xhline{1.5\arrayrulewidth}
\multirow{6}*{\rotatebox{90}{Wav2vec2-base-960h}} 
& G1 & 0.4961 & 0.3039 & 0.9982 & 0.0014 & 0.0014 \\ 
& G2 & 0.4886 & 0.2830 & 0.9989 & 0.0015 & 0.0015 \\ 
& G3 & 0.5564 & 0.2940 & 0.9760 & 0.0439 & 0.0439 \\ 
& G4 & 0.4812 & 0.2888 & 0.9999 & 0.0002 & 0.0002 \\ 
& G5 & 0.4922 & 0.2883 & 0.9999 & 0.0000 & 1.8E0-05 \\
& G6 & 0.5122 & 0.2791 & 0.9965 & 0.0017 & 0.0017 \\ 
\Xhline{1.0\arrayrulewidth}
\multirow{6}*{\rotatebox{90}{Wav2vec2-large-960h}} 
& G1 & 0.5126 & 0.2968 & 0.9892 & 0.0094 & 0.0094 \\
& G2 & 0.5109 & 0.2911 & 0.9987 & 0.0003 & 1.0003 \\ 
& G3 & 0.4762 & 0.2984 & 0.9583 & 0.0162 & 0.0162 \\ 
& G4 & 0.4933 & 0.2836 & 0.9997 & 0.0008 & 0.0008 \\ 
& G5 & 0.5026 & 0.2907 & 0.9999 & 0.0001 & 0.0001 \\ 
& G6 & 0.5229 & 0.2926 & 0.9842 & 0.0002 & 0.0002\\
\Xhline{1.0\arrayrulewidth}
\multirow{6}*{\rotatebox{90}{Hubert-base-ls960}} 
& G1 & 0.5276 & 0.2996 & 0.9986 & 0.0100 & 0.0100 \\ 
& G2 & 0.5035 & 0.2829 & 0.9968 & 0.0006 & 0.0006 \\ 
& G3 & 0.5139 & 0.2961 & 0.9779 & 0.0379 & 0.0279 \\ 
& G4 & 0.5129 & 0.2902 & 0.9986 & 0.0001 & 0.0001 \\ 
& G5 & 0.5009 & 0.2879 & 1.0000 & 0.0001 & 0.0001 \\
& G6 & 0.4934 & 0.3022 & 0.9924 & 0.0021 & 0.0021 \\ 
\Xhline{1.5\arrayrulewidth}
\end{tabular}
\label{table:wave2_table3}
\vspace{-0.4cm}
\end{table}

\textbf{2. Fine-Trued Models}

To assess the impact of dataset-specific fine-tuning, we retrained three pretrained transformer-based models—Wav2Vec2-base-960h (Model A), Wav2Vec2-large-960h (Model B), and HuBERT-base-ls960 (Model C)—on the HSAD dataset. Training was conducted using 20 epochs with the Adam optimizer, cosine learning rate schedule, and early stopping. Input audio was standardized to 128-bin log-Mel spectrograms with 25ms frame length and 10ms hop size. All models were trained on an NVIDIA A100 GPU using a batch size of 3.

As shown in the updated confusion matrices, each model exhibited substantial improvements in classification performance following fine-tuning. Model A (fine-tuned Wav2Vec2-base-960h) achieved an accuracy of 97.94\%, with balanced true positive and true negative rates (TP = 1681, TN = 1647). Model B (fine-tuned Wav2Vec2-large-960h) slightly lagged with 97.26\% accuracy, while Model C (fine-tuned HuBERT-base-ls960) achieved 96.76\%, still representing a dramatic leap over their original baseline accuracies (20.06\%, 3.52\%, and 1.30\%, respectively).

\begin{figure}[t]
    \centering
    \includegraphics[width=0.5\linewidth]{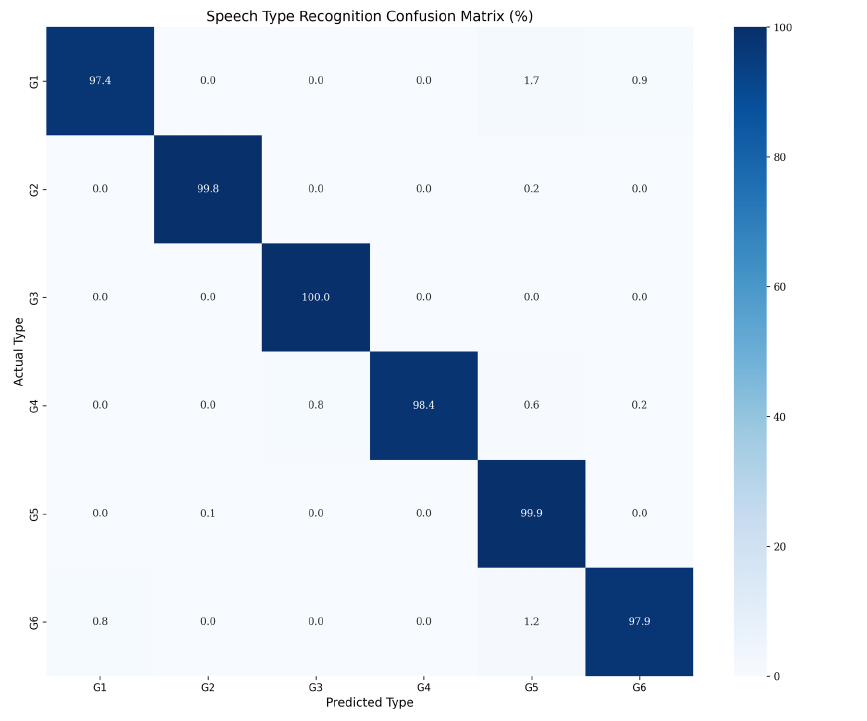}
    \caption{Confusion matrix – Wav2vec2-base-960h fine-tuned (Model A)}
\end{figure}
\begin{figure}[t]
    \centering
    \includegraphics[width=0.5\linewidth]{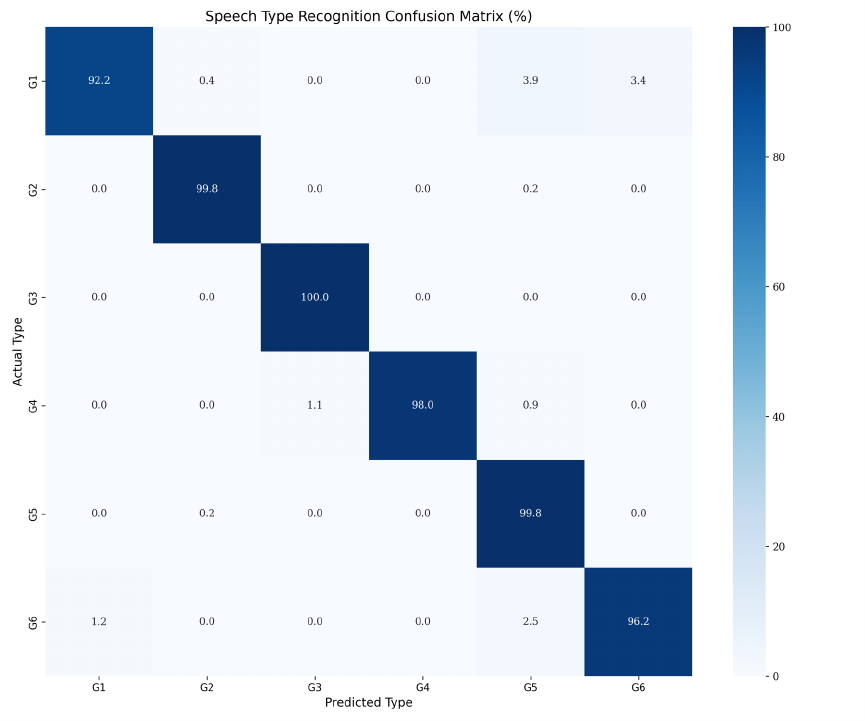}
    \caption{Confusion matrix – Wav2vec2-large-960h fine-tuned (Model B)}
\end{figure}

\begin{figure}[t]
    \centering
    \includegraphics[width=0.5\linewidth]{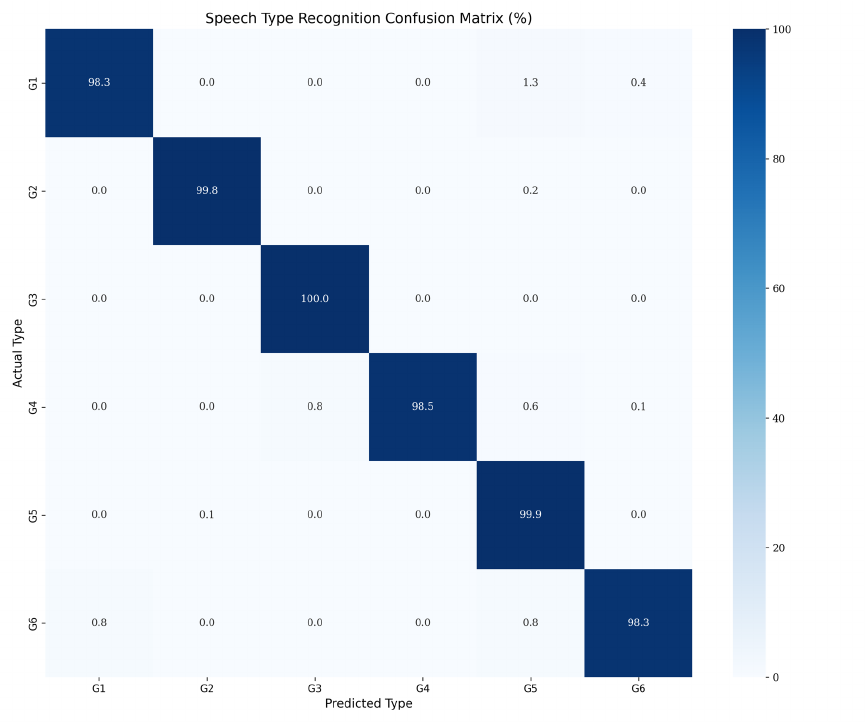}
    \caption{Confusion matrix – Hubert-base-ls960 fine-tuned (Model C)}
\end{figure}
Across all three fine-tuned models, false positives and false negatives were significantly reduced. As shown in Fig.~\ref{fig:all_comb}, Model A reduced human speech misclassifications by over 90\%, while Model B minimized synthetic speech false negatives by approximately 88\%. Model C, despite a slightly lower accuracy, displayed robust generalization to hybrid and ambiguous samples. All three models achieved F1-scores above 96\%, marking a vast improvement from the high-entropy, low-certainty decisions observed in their pretrained counterparts.

\begin{figure}[t]
    \centering
    \includegraphics[width=\linewidth]{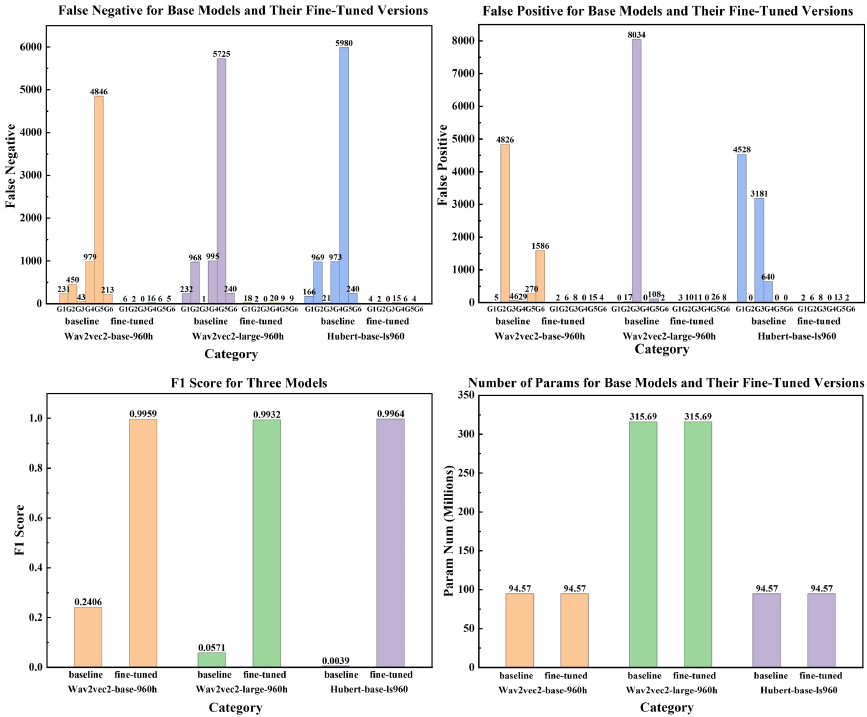}
    \caption{Comparison of false positives, false negatives, F1-score, and parameter count across baseline and fine-tuned Wav2vec2-base-960h, Wav2vec2-large-960h, and Hubert-base-ls960 fine-tuned}
    \label{fig:all_comb}
\end{figure}

\subsection{Discussion}

This study introduces a systematically constructed Hybrid and Spoofed Audio Dataset (HSAD) designed to expose the limitations of current state-of-the-art anti-spoofing systems under complex and adversarial conditions. Unlike existing corpora such as ASVspoof 2019, which focus predominantly on binary classification and clean synthetic speech, our dataset reflects real-world threat vectors through four spoofing classes: genuine human speech, AI-cloned speech, fully AI-generated speech, and various hybrid compositions combining human and synthetic audio segments. Additionally, we inject practical distortions such as environmental noise, codec compression, and channel degradation to simulate mobile and cross-platform transmission conditions.

Empirical results from this work highlight a significant performance gap between models trained on homogeneous public datasets and their effectiveness when deployed in adversarial environments. Although the MIT/AudioSet transformer model achieved over 93\% accuracy in aggregate, it misclassified all genuine human samples as spoofed, revealing a critical flaw in model calibration. The MattyB95 baseline, despite being fine-tuned on the ASVspoof 2019 dataset, showed improved class separability but struggled with mixed-source inputs due to overlapping reliability distributions.

In contrast, fine-tuned models trained directly on HSAD achieved substantially higher performance across all metrics. With accuracy exceeding 97\% and F1-scores approaching 99\%, these models reduced false positive rates for human speech by over 70\% and false negatives for synthetic speech by over 90\%. These results not only validate the advantage of dataset-specific fine-tuning, but also emphasize the importance of including hybridized spoof structures during training to account for increasingly subtle and compositional spoofing techniques.






\subsection{Limitations and Future Works}
\subsubsection{Limitations}

Despite the substantial contributions of this work, several limitations must be acknowledged to contextualize the findings:

\textbf{1. Spoofing Coverage.} The proposed HSAD dataset currently focuses on two primary spoofing paradigms: text-to-speech (TTS) synthesis and AI-based voice cloning. It does not yet incorporate advanced attack vectors such as voice conversion (VC), prosody manipulation, GAN-generated speech, or adversarial examples. This restricts the coverage of attack modalities and may limit model generalization to unseen spoofing types.

\textbf{2. Environmental Realism.} Although HSAD introduces artificial degradation such as additive noise, codec compression, and channel filtering, all recordings originate from controlled environments. Real-world variability—including spontaneous speech, overlapping speakers, far-field microphones, and transmission distortions—remains underrepresented, potentially biasing evaluation.

\textbf{3. Dataset Scale.} The current speaker pool is limited in both size and linguistic diversity. While balanced across age and gender, it may not represent broader phonetic variability or demographic factors, which can affect spoof detection in diverse real-world settings.

\textbf{4. Annotation Granularity.} Hybrid audio samples are currently labeled at the utterance level, without frame- or segment-level annotation of spoofed regions. This limits the dataset’s utility for training models capable of fine-grained spoof localization or real-time detection.

\subsubsection{Future Work}

Building upon the current framework, we identify several future directions to enhance the dataset and improve model robustness:

\textbf{1. Expansion of Spoof Modalities.} Future iterations of HSAD will include broader spoofing types such as VC-based attacks, GAN-based speech generation, multilingual synthesis, and adversarial examples to create a more comprehensive benchmark.

\textbf{2. Real-World Data Collection.} To better emulate deployment conditions, we plan to incorporate in-the-wild speech captured from smartphones, IoT devices, and telephony systems across diverse acoustic environments, including noisy, reverberant, and far-field scenarios.

\textbf{3. Segment-Level Annotation.} We will introduce precise time-stamped annotations indicating the onset and offset of synthetic segments within hybrid utterances. This will support fine-grained spoof localization and improve training for segment-wise detection models.

\textbf{4. Cross-Corpus Transferability.} We will evaluate fine-tuned models on additional corpora such as ADD2023-PF, Half-and-Half (HAD), and multilingual spoof detection datasets to assess generalization and cross-domain performance.

\textbf{5. Explainability and Robustness.} Future work will explore interpretable AI techniques such as saliency mapping and attention visualization to understand spoof detection decisions. In addition, robustness against adversarial perturbations and distribution shifts will be systematically evaluated.

\noindent Together, these directions aim to advance the development of anti-spoofing systems that are not only accurate and generalizable, but also explainable and deployable in adversarial and heterogeneous environments.

%% file: 06-Conclusion.tex
\section{Conclusion}

This work introduced the Hybrid Spoofed Audio Dataset (HSAD), a novel benchmark designed to support robust and generalizable detection of complex and adversarial audio spoofing threats. Unlike prior corpora restricted to binary spoof labels, HSAD features fine-grained spoof categories—including AI-cloned, zero-shot synthesized, and hybrid compositions—generated using Tacotron2-based voice cloning, ElevenLabs zero-shot TTS, and intra-speaker splicing. Real-world acoustic conditions were simulated through codec compression, additive noise, and spectral filtering.

To evaluate model performance, we tested five transformer-based architectures: two Audio Spectrogram Transformer (AST) variants (MIT-AST and MattyB95-AST) and three self-supervised speech models (Wav2Vec2-base, Wav2Vec2-large, and HuBERT-base). Our results revealed critical insights: pretrained models struggled to generalize when exposed to hybrid and cross-domain spoof patterns, with confusion particularly evident between cloned and AI-generated categories. The MattyB95-AST model, although fine-tuned on ASVspoof 2019, misclassified genuine human samples and exhibited low confidence on mixed utterances.

Upon fine-tuning all models on HSAD, substantial gains were achieved. AST models surpassed 97\% classification accuracy with F1-scores approaching 99\%, and Wav2Vec2-large and HuBERT-base demonstrated competitive spoof discrimination after domain adaptation. Confusion matrices showed clear improvements in inter-class separation, and performance was robust under spectrally degraded or hybridized conditions. These findings confirm the importance of dataset-specific fine-tuning and the inadequacy of legacy benchmarks for evaluating real-world spoof detection.

This study also highlighted the utility of integrating general-purpose and ASR-pretrained models for spoof detection. While ASTs benefited from visual transfer learning, self-supervised speech models like Wav2Vec2 and HuBERT offered strong representational priors from unlabeled speech corpora, achieving high performance with modest fine-tuning.

Future research will expand HSAD to include multilingual speakers, adversarial attacks, voice conversion (VC), and fine-grained segment-level spoof annotations. Real-world recordings from telephony and conferencing platforms will further enhance environmental realism. Additionally, explainable AI (XAI) methods will be explored to interpret spoof attribution at the token or frame level.

\noindent In conclusion, HSAD establishes a comprehensive and realistic foundation for next-generation spoof detection research. It demonstrates that modern transformer models, when properly adapted, can achieve high robustness against complex spoofing threats—paving the way toward secure, interpretable, and real-world deployable ASV systems.

%% file: 07-Acknowledgement.tex
\vspace{-0.3cm}
\section{Acknowledgement}
\vspace{-0.3cm}
This work has been submitted to the IEEE for possible publication. Copyright may be transferred without notice, after which this version may no longer be accessible.